\documentclass[10pt,twoside]{IEEEtran}

\usepackage[latin1]{inputenc}
\usepackage{amssymb}
\usepackage{amsmath}
\usepackage{amsthm}
\usepackage{mathrsfs}
\usepackage{graphicx}
\usepackage{array}
\usepackage{tabularx}
\usepackage{subfigure}
\usepackage{caption}
\usepackage{xcolor}
\usepackage{cite}
\usepackage{psfrag}
\usepackage[keeplastbox,nospread]{flushend}

\begin{document}

\title{Comments on ``Finite-SNR Diversity-Multiplexing Tradeoff for
Network Coded Cooperative OFDMA Systems''}

\author{Thiago Henrique Ton, João Luiz Rebelatto and Richard Demo Souza}

\maketitle

\begin{abstract}
In the paper ``Finite-SNR Diversity-Multiplexing Tradeoff for Network Coded Cooperative OFDMA Systems'', the authors adopt a recently proposed subcarrier allocation algorithm aiming at providing optimal frequency diversity to an OFDMA-based network with network-coded cooperation (NCC), where $P$ source nodes transmit independent information to a common destination with the help of $M$ relay nodes, in a frequency-selective channel with $L$ coherence blocks. However, the analysis of the so-called NCC-OFDMA scheme proposed in the aforementioned paper assumes that the subcarrier allocation algorithm is capable of providing full frequency diversity to all the links in the network, which generally does not hold in practice since the allocation is optimal only from the perspective of a single destination. Therefore, in this note we reevaluate the outage probability of the NCC-OFDMA under a more realistic scenario where only the messages addressed to the destination are benefited with the full frequency diversity provided by the subcarrier allocation algorithm, showing that the diversity order is decreased from $L(M+1)$ to $M+L$.
\end{abstract}


\section{Introduction}


In~\cite{heidarpour.17.twc}, Heidarpour {\it et.al.} adopt the maximum constraint $\mathcal{K}_{1,K}$-matching allocation (MCMA) algorithm from~\cite{bobai.11.hmatching} to allocate the subcarriers and increase the diversity order of an OFDMA network where $P$ sources nodes transmit to a single destination and are assisted by a set of $M$ dedicated relay nodes, which employ the non-binary network-coded cooperation (NCC) technique. According to~\cite{heidarpour.17.twc}, the proposed NCC-OFDMA scheme is capable of achieving diversity order $L(M+1)$ in a frequency selective channel with $L$ coherence blocks. However, the authors in~\cite{heidarpour.17.twc} consider an optimistic assumption in many practical setups that the allocation algorithm from~\cite{bobai.11.hmatching} is capable of providing full frequency diversity at the same time throughout all the links in the cooperative network.


Following~\cite[Theorem 2]{bobai.11.hmatching}, if the destination of an OFDMA system with $L$ coherence blocks allocates $K$ subcarriers to each source node according to the MCMA algorithm, then the outage probability of a given frame experienced at the destination is
\begin{equation}\label{eq:outage_mcma}
P_D = \left[P_{{out}_{sub}}(K)\right]^L,
\end{equation}
where $P_{{out}_{sub}}(K)$ is the single-carrier outage probability as presented in~\cite[(10-12)]{heidarpour.17.twc}. 

When calculating~\cite[(14)]{heidarpour.17.twc}, the authors consider that the outage probability of multiple source-to-relays links are obtained according to~\eqref{eq:outage_mcma}. However, the result from~\eqref{eq:outage_mcma} was obtained in~\cite{bobai.11.hmatching} considering a single-hop multiple access non-cooperative network, where multiple users independently transmit their own information to a common destination. That is, all the transmissions are addressed uniquely to a single destination, which is responsible for properly distributing/allocating the subcarriers among the source nodes. 
In a network with multiple receivers subject to independent fading, the diversity order $L$ from~\eqref{eq:outage_mcma} can be obtained only at the receiver node where the subcarrier allocation is assigned. The other receivers would not experience any frequency diversity, as the subcarrier allocation would appear random to them, as discussed in details in what follows.

\begin{figure}[!t]
\begin{center}
\psfrag{u1}[cr][cr]{$u_1$}
\psfrag{u2}[cr][cr]{$u_2$}
\psfrag{u3}[cr][cr]{$u_3$}
\psfrag{u4}[cr][cr]{$u_4$}
\psfrag{s1}[cl][cl]{$s_1$}
\psfrag{s2}[cl][cl]{$s_2$}
\psfrag{s3}[cl][cl]{$s_3$}
\psfrag{s4}[cl][cl]{$s_4$}
\psfrag{s5}[cl][cl]{$s_5$}
\psfrag{s6}[cl][cl]{$s_6$}
\psfrag{s7}[cl][cl]{$s_7$}
\psfrag{s8}[cl][cl]{$s_8$}
\psfrag{a}[c][c]{{\footnotesize Users}}
\psfrag{b}[c][c]{{\footnotesize Subcarriers}}
\subfigure[\label{fig:graphD}Users-Destination]{\includegraphics[width=0.2\textwidth]{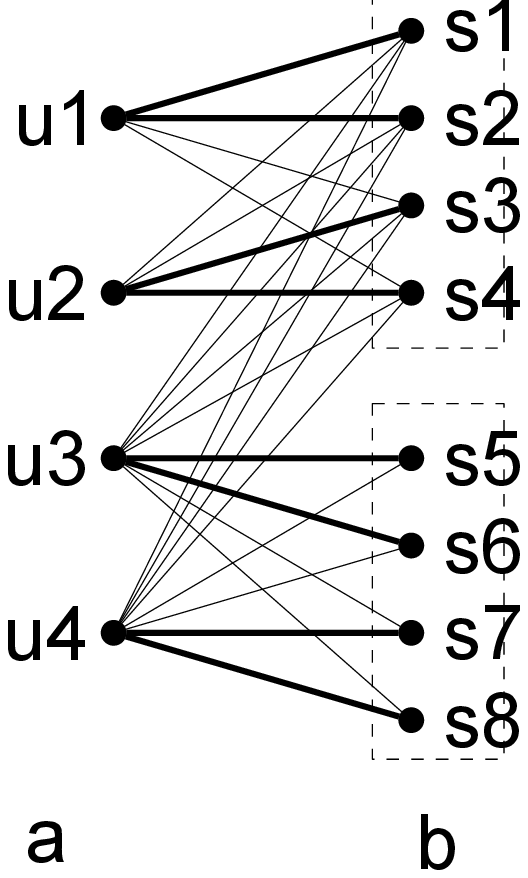}}
\qquad
\subfigure[\label{fig:graphR}Users-Relay]{\includegraphics[width=0.2\textwidth]{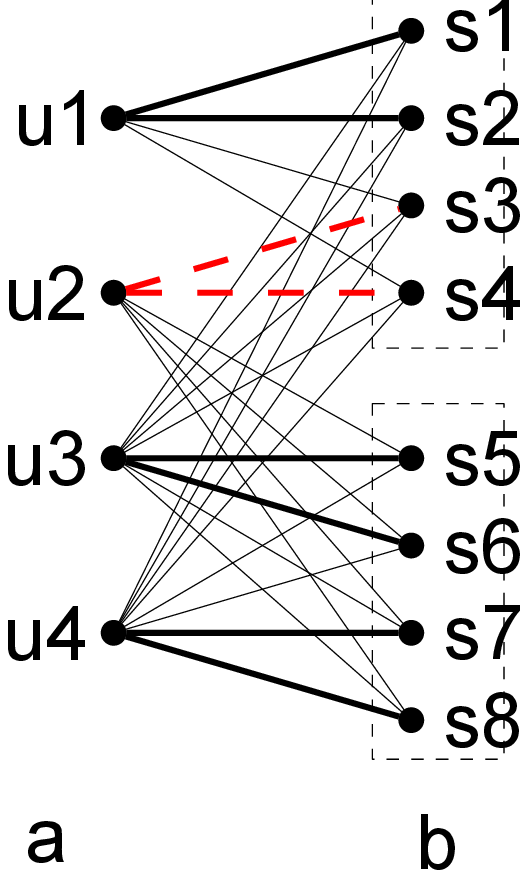}}
\caption{An example of the maximum constraint $\mathcal{K}_{1,k}$-matching from the sources to the destination node, which is denoted by thick lines in~\ref{fig:graphD}. In Fig.~\ref{fig:graphR}, it is presented the bipartite graph from the relay perspective, where the dashed-red edges represent the edges that belong to $\varepsilon$ but not to $\varepsilon_{\tt R}$, which means that $u_2$ will be in outage at the relay. The dashed box denotes a coherence bandwidth, which in the example is composed of $N_c = 4$ subcarriers.}
\label{fig:example}
\end{center}
\end{figure}

Following~\cite{bobai.11.hmatching}, let $\mathcal{G}(\mathcal{U}\cup\mathcal{S},\varepsilon)$ be the bipartite graph that connects the users from the set $\mathcal{U}$ to subcarriers from the set $\mathcal{S}$, as illustrated in Fig.~\ref{fig:graphD}. The vertices $u_m$ and $s_n$ are connected through edge $e_{mn}$ as long as subcarrier $s_n$ is not in outage for user $u_m$. The set of all edges (connection between users and subcarriers) is represented by $\varepsilon$, while $\mathcal{N}(u_m)$ contains all the subcarriers that are not in outage for user $u_m$.

According to~\cite[Lemma 1]{bobai.11.hmatching}, one can see that the bipartite graph $\mathcal{G}(\mathcal{U}\cup\mathcal{S},\varepsilon)$ presented in Fig.~\ref{fig:graphD} has a maximum constraint $\mathcal{K}_{1,k}$-matching $\mathcal{M}^c_{\mathcal{K}_{1,k}}$ that saturates every vertex in $\mathcal{U}$, since the condition $|\mathcal{N}(\mathcal{X})| \geq K|\mathcal{X}|$ is met $\forall \mathcal{X} \subseteq \mathcal{U}$. In the example from Fig.~\ref{fig:graphD}, $\mathcal{M}^c_{\mathcal{K}_{1,k}} = \{u_1s_1, u_1s_2,u_2s_3, u_2s_4,u_3s_5, u_3s_6,u_4s_7, u_4s_8\}$ (represented by the thick lines). Thus, there will be no user in outage at the destination node.

\subsection{The Relays do not Achieve the Full Diversity Order from~\cite[Theorem 2]{bobai.11.hmatching}}

Let us now evaluate the bipartite graph in this example from a perspective of a receiver other than the destination. Without loss of generality, we consider the scenario from~\cite{heidarpour.17.twc} with $M=1$ relay, which, besides the destination node, also attempts to recover the messages from the users. A snapshot of the bipartite graph from the users to the relay is represented by $\mathcal{G}(\mathcal{U}\cup\mathcal{S},\varepsilon_{\tt R})$ and illustrated in Fig.~\ref{fig:graphR}.

In the example, $\mathcal{M}^c_{\mathcal{K}_{1,k}}$ still saturates users $u_1$, $u_3$ and $u_4$ at the relay node, since $\mathcal{N}_{\tt R}(u_m) = \mathcal{N}(u_m)$ for $m = 1,3,4$. That is, the set of subcarriers not in outage for such users are the same to both $\mathcal{G}(\mathcal{U}\cup\mathcal{S},\varepsilon)$ and $\mathcal{G}(\mathcal{U}\cup\mathcal{S},\varepsilon_{\tt R})$. However, the same cannot be said about $u_2$, whose subcarriers not in outage are now given by  $\mathcal{N}_{\tt R}(u_2) = \{s_5, s_6, s_7, s_8\}$. Thus, since the destination has allocated subcarriers $s_3$ and $s_4$ to $u_2$ (which are represented by the red-dashed line in Fig.~\ref{fig:graphR}), such user will be in outage at the relay node. Even though the condition $|\mathcal{N}_{\tt R}(\mathcal{X})| \geq k|\mathcal{X}|$ is also met $\forall \mathcal{X} \subseteq \mathcal{U}$ at the relay, it is not subject to the results from~\cite[Theorem 2]{bobai.11.hmatching}. Consequently, the links between the users and the relay do not achieve diversity order $L$. This is due to the subcarrier allocation being based exclusively on $\varepsilon$, not depending on $\varepsilon_{\tt R}$. Therefore, the subcarrier allocation at the users, as defined by the destination, seems random at any other receiving node, as the relay.

\subsection{Diversity Order Achieved by the Relays}

According to~\cite[Theorem 2]{bobai.11.hmatching}, an user will be in outage at the destination node responsible for generating $\mathcal{M}^c_{\mathcal{K}_{1,k}}$ only when such user corresponds to an isolated vertex in the maximum constraint $\mathcal{K}_{1,k}$-matching. Such an event occurs when all the edges in all the $L$ coherence blocks do not exist, whose probability decays with SNR$^{-L}$ and makes the diversity order experienced at the destination to be $L$, as presented in~\eqref{eq:outage_mcma}. However, from Fig.~\ref{fig:graphR}, we can see that, differently from the destination node, a single outage event, whose probability is given by $P_{{out}_{sub}}(K)$, has led $u_2$ to be in outage at the relay, concluding that the relay experiences diversity order 1.

Alternatively, assume without loss of generality that $M=N_c$ as in~\cite[Remark 1]{bobai.11.hmatching}. Then, given the random subcarrier allocation seen by the relay, the outage probability experienced at the relay cannot be smaller than the interleaved subcarrier allocation, which, for $k=K$, decays with SNR$^{-1}$~\cite[Remark 1]{bobai.11.hmatching}, leading to an unitary diversity order.

Therefore, considering the particular setup in~\cite{heidarpour.17.twc}, where in the broadcast phase the $P$ sources transmit to the $M$ relays and to the single destination through $K_1$ subcarriers, from the relays point-of-view, the subcarrier allocation performed by the destination is the same as if the subcarriers were randomly distributed among the source nodes, since the channels are assumed to be i.i.d. across time and space. This leads the outage probability of a single frame in the source-to-relay link to be
 $P_{{out}_{sub}^{SR}}(K_1)$~\cite[(11)]{heidarpour.17.twc} instead of~\eqref{eq:outage_mcma}, without experiencing the diversity order $L$ provided by the subcarrier allocation.


In this note we reevaluate the performance of NCC-OFDMA scheme from~\cite{heidarpour.17.twc} under the above realistic scenario where only the messages addressed to the destination are benefited with the full frequency diversity provided by the subcarrier allocation algorithm from~\cite{bobai.11.hmatching}. 

\section{Proposed Outage Probability Analysis} \label{sec:outage}

From~\cite[(13)]{heidarpour.17.twc}, the overall outage probability of the NCC-OFDMA scheme is
\begin{equation}
P_{out} = \sum_{m=0}^{M} P_{out,m} P_m,
\end{equation}
where the term $P_m$ is still obtained as~\cite[(15)]{heidarpour.17.twc}, since does not depend on the outage probability in the source-to-relay links. However, the term $P_R$ from $P_{out,m}$ (probability that any relay successfully decodes all $P$ packets received during the broadcasting phase, see~\cite[(14)]{heidarpour.17.twc}) is now
\begin{equation} \label{eq:p_r}
P_R = \left[1-P_{{out}^{SR}_{sub}}(K_1)\right]^P,
 \end{equation}where $P_{{out}^{SR}_{sub}}(K_1)$ is given by~\cite[(11)]{heidarpour.17.twc}, since the source-to-relay links do not experience frequency diversity due to the subcarrier allocation assigned at the destination node.
\begin{figure}[!t]
\centering{}
\includegraphics[width=0.45\textwidth]{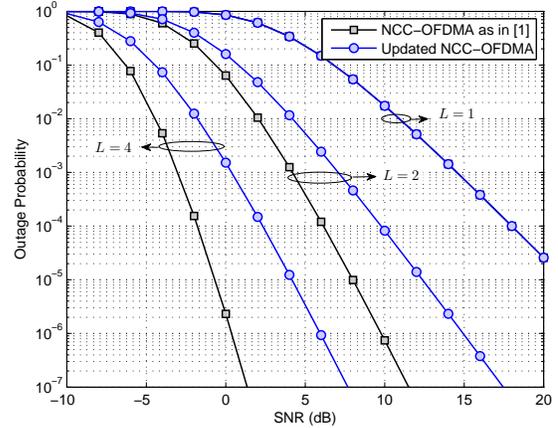}
\caption{Outage probability of the NCC-OFDMA scheme as proposed in~\cite{heidarpour.17.twc} and as updated in this note, for different values of $L$ (with $P=M=2$, transmission rate $R_0=1$ bps/Hz).}
\label{fig:outage_L124}
\end{figure}

For illustrative purposes, if we consider a particular scenario where $P=M$, it turns out that the outage probability of the NCC-OFDMA scheme becomes:
\begin{equation} \label{eq:outage_ncc}
P_{out} = [M-1]^M P_{{out}_{sub}}^{M+L} + \textsf{O}\left(P_{{out}_{sub}}^{M+L}\right),
\end{equation}
where $\textsf{O}(x)$ is the higher order infinitesimal of $x$. The result in~\eqref{eq:outage_ncc} is obtained after replacing~\eqref{eq:p_r} in~\cite[(16)]{heidarpour.17.twc} and then expanding it, noting that the most relevant term, at high SNR, is obtained when $m=M$, $j=0$ and $i=P-1$.

From~\eqref{eq:outage_ncc} we can see that the diversity order achieved by the NCC-OFDMA scheme is equal to $M+L$, rather than $L(M+1)$ as stated in~\cite{heidarpour.17.twc}. It can be shown that the same results are obtained for $P \neq M$. It is also worth noting that the result from~\eqref{eq:outage_ncc} also impacts on the diversity gain of the asymptotical DMT presented in~\cite[(27)]{heidarpour.17.twc}, which now is bounded by $M+L$, as well as on the finite-SNR DMT from~\cite[(40)]{heidarpour.17.twc} and~\cite[(52)]{heidarpour.17.twc}.


The difference between the outage probabilities present in~\cite{heidarpour.17.twc} and~\eqref{eq:outage_ncc} is illustrated in Fig.~\ref{fig:outage_L124} for some values of $L$, and considering Rayleigh fading. One can see that the updated outage probability of the NCC-OFDMA scheme significantly differs from the analysis in~\cite{heidarpour.17.twc}, and that such difference increases with $L$.

\bibliographystyle{IEEEtran}
\bibliography{IEEEabrv,references}

\end{document}